\documentclass[a4paper]{VisionStyle}
\usepackage{epsfig}

\begin{document}

\title{A full orbit XMM-Newton observation of PKS 2155-304}

\author{Laura Maraschi\inst{1} \and
Fabrizio Tavecchio\inst{1} \and Ilaria Cagnoni\inst{2} \and You-Hong Zhang\inst{2} \and
Lucio Chiappetti\inst{3} \and Aldo Treves\inst{2} \and Annalisa Celotti\inst{4} \and
Luigi Costamante\inst{5} \and Rick Edelson\inst{6,7} \and Giovanni Fossati\inst{8} \and
Gabriele Ghisellini\inst{5} \and Elena Pian\inst{9} \and Steve Sembay\inst{7} \and 
Gianpiero Tagliaferri\inst{5} \and
C. Megan Urry\inst{10}
} 

\institute{Osservatorio Astronomico di Brera, Via Brera 28 I-20121
Milano, Italy \and Dipartmento di Scienze, Universit\`a dell'Insubria,
via Valleggio 11 Como, I-22100, Italy \and Instituto di Fisica Cosmica
G. Occhialini, via Bassini 15, Milano, I-20133, Italy \and SISSA/ISAS,
via Beirut 2-4 Trieste, I-34014, Italy \and Osservatorio Astronomico di
Brera, via Bianchi 46 Merate, I-23807, Italy \and Astronomy Department,
University of California, Los Angeles, CA 90095-1562, USA \and X-Ray
Astronomy Group, University of Leicester, Leicester LE1 7RH, UK \and
Dept. of Physics and Astronomy, MS 108, 6100 Main street, Houston, TX,
77005 USA \and Osservatorio Astronomico di Trieste, via G. B. Tiepolo 11,
Trieste, I-34131, Italy \and Yale Center for Astronomy \& Astrophysics,
PO Box 208121, New Haven CT 06520-8121, USA}

\maketitle 

\begin{abstract}

XMM observed the BL Lac PKS 2155-304 for a full orbit ($\sim$150 ksec) on
2000 November 19-21. Preliminary results on the temporal and spectral
analysis of data from the EPIC PN camera and Optical Monitor are
presented. The variability  amplitude depends systematically on energy, 
however the slopes of the structure functions of the light-curves in
different bands do not appear to be significantly different.
No evidence of time lags is found by cross correlating the light-curves 
in different bands.
\end{abstract}

\section{Introduction}

Emission from Blazars is dominated by the non-thermal continuum produced
in a relativistic jet pointing close to the line of sight (e.g. Urry \&
Padovani 1995).  Thanks to the effects of relativistic beaming the jet
flux can be enhanced by several orders of magnitude and its variability
can therefore be studied observationally down to short timscales.
Blazars hold crucial information on the physical processes taking place
in relativistic jets.  The ``double humped'' shape of the Spectral Energy
Distribution (SED) is generally interpreted as due to synchrotron and
Inverse Compton emission from a population of relativistic electrons in
the jet.  In fact from modelling the shape of the SED and the correlated
variability of the two components it is possible to constrain the basic
physical quantities of the jet (e.g. Sikora \& Madejski 2001; Tavecchio
et al. 1998). In particular, variability allows one to explore the
acceleration and cooling processes acting on electrons. On this regard,
especially valuable are observations in X-rays, which are produced by
high-energy, fast-evolving particles.

Thanks to the {\em ASCA} and {\em Beppo-SAX} satellites, time lags
between variability at different X-ray energies were discovered in PKS
2155-304 and Mkn 421. In the case of PKS 2155-304 lags were always of the
same sign, soft X-rays lagging the harder ones, while both signs were
found in Mkn 421.  The simplest interpretation of these findings is that
the emission from different X-ray energy bands is due to the same
population of electrons.  Within this picture the observed lags
correspond to the time needed by the electrons to cool (soft lag) or
accelerate (hard lag) and emit radiation in another energy band.

The main limitation to the {\em ASCA} and {\em Beppo-SAX} studies is
related to the low orbit of such satellites: the periodic earth
blockage prevents a continuous sampling of flux behaviuours, making
temporal analysis more complicated and less reliable.  The advent of
the new generation satellites, such as {\em XMM-Newton} and {\em
Chandra}, thanks to their high altitude and elliptical orbits, opens a
new era in variability studies.  In fact they offer continuous target
visibility for up to about 40 hours compared with only $\sim 90$
minutes from the low orbit satellites. In particular the big
improvement offered by {\em XMM-Newton} is due to the combination of
such a long continuous coverage of the target coupled with an
unprecedently large effective area. Another striking feature of {\em
XMM-Newton} is the possibility offered by the Optical Monitor to
explore simultaneously the optical-UV bands.

With the goal of characterizing in a better way the variability we
obtained a $\sim 150$~ks {\em XMM-Newton} observation of one of the X-ray
and UV brightest Blazars: PKS 2155-304.

PKS 2155-304 ($z=0.116$) is a well known BL Lac object (a subclass class
of blazars with an almost featureless optical spectrum), intensively
monitored in the past years. Limiting quotations to results based on {\em
ASCA}, {\em BeppoSAX}, {\em CHANDRA} or {\em XMM-Newton}, a minimal list
of references is the following: Urry et al. 1997 (reporting simultaneous
{\em ASCA}, {\em EUVE} and {\em IUE} lightcurves); Tanihata et al. 2001
(continuous 10-days {\em ASCA} observation); Kataoka et al. 2001 (summary
of {\em ASCA} and {\em Rossi-XTE} lightcurves); Zhang et al. 2002
(summary of {\em BeppoSAX} lightcurves); Nicastro et al. 2002 (the first
spectrum of PKS 2155-304 obtained with the {\em Chandra}
gratings). Edelson et al. (2001) report on a 100 ks-long {\em XMM-Newton}
observation secured on May 30 2000 during the Performance
Verification/Guaranteed Time Observation phase.\\

We present the results of a preliminary analysis of a long XMM-Newton
pointing of PKS 2155-304 on Nov 19-21 2000.  Due to problems in the
satellite data analysis pipeline, the preprocessed data became available
to us only recently.  We present in this paper a preliminary analysis of
the data of the EPIC PN camera and of the Optical Monitor (OM) only.
EPIC MOS1 and MOS2 and RGS data will be discussed elsewhere (however see
Brinkmann et al. in these proceedings).

\section{Results}

\subsection{Data Analysis}

PKS 2155-304 was observed during two uninterrupted pointings
from Nov 19 (18:47:00 UT) to Nov 20 (10:47:49 UT) and from Nov 20
(13:01:39 UT) to Nov 21 (05:17:28 UT), respectively. The gap between the
two parts is $\sim $2 hrs.\\

In order to avoid pile-up effects due to the large flux of the source,
the PN camera was operated in the Small Window mode and the two MOS
cameras in Timing Mode (MOS1) and Small Window Mode (MOS2). 
%We limit the present discussion to PN and OM data only.  
The data, processed with the
standard pipeline, were analyzed with the {\it XMM-Newton Science
Analysis System} (version 5.2.0).\\

\begin{figure}[ht]
  \begin{center}
    \epsfig{file=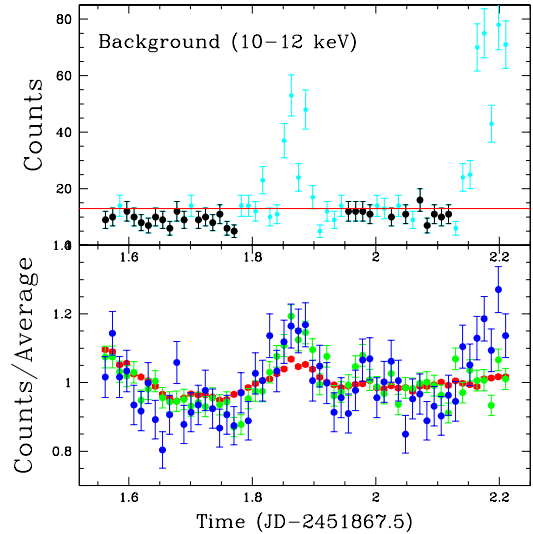, width=8cm}
  \end{center}
\caption{Background (top) and  normalized EPIC PN  (bottom) lightcurves 
of the second part of the observation. 
In the bottom panel red circles correspond to the 0.1--2 keV energy band,
green the 2-4 keV and blue the 4-10 keV band.
The red line in the top panel represents our cut: the cyan points represent
the intervals of time we excluded.}
\label{lmaraschi-C2_fig:unclean}
\end{figure}

A well known problem in EPIC  data analysis is the background
stability (e.g. Katayama et al. 2002).  The background spectrum is
dominated by particle events above 5 keV (e.g. Katayama et al. 2002).
PKS 2155-304 emits most of its photons in the soft X-ray band
so we were
able to assess the stability of the background using a 1000~s binned   light
curve from events between 10 and 12 keV in a circular region
 ($r=100^{\prime \prime}$) far from the source position  
(Figure~\ref{lmaraschi-C2_fig:unclean} top).
The first part of the observation has a stable background (with a mean of
10 counts per bin), while the second part shows changes $\sim 8$ times
larger than the quiescent level of $\sim 9.6$ counts per bin.  We
conservatively excluded all the time intervals of the second part of the
observation with background (10-12~keV) counts per bin larger than 13
(red line in Figure~\ref{lmaraschi-C2_fig:unclean}).

\begin{figure}[ht]
  \begin{center}
    \epsfig{file=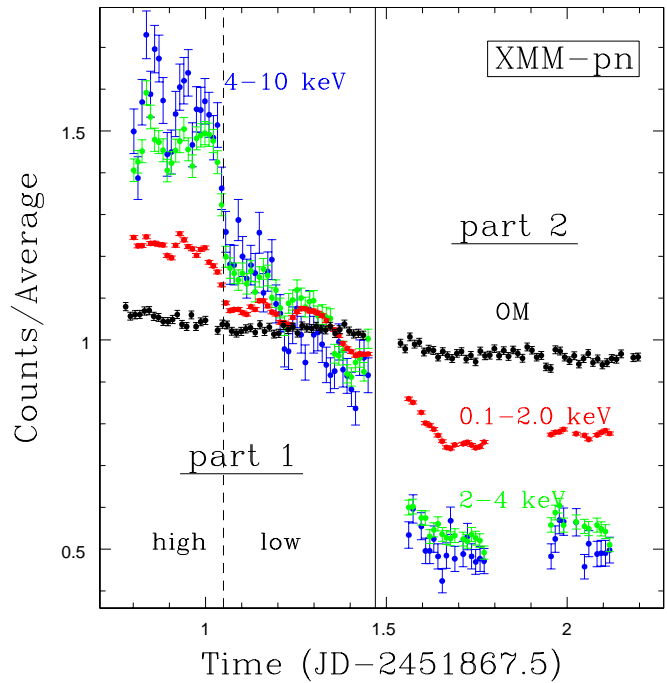, width=10cm}
  \end{center}
\caption{OM and EPIC PN lightcurves normalized to the mean. 
Black symbols represent OM data, red circles correspond to the 0.1--2 keV 
energy band, green the 2-4 keV and blue the 4-10 keV band.}
\label{lmaraschi-C2_fig:lc}
\end{figure}

We extracted light curves in 3 energy bands, 0.1--2 keV, 2--4 keV and 4--10
keV with a bin size of 1000~s in a circular region with $r = 40^{\prime
\prime}$ centered on the source position.  Due to a visible trend of
increasing source flux after the end of a consistent ($\geq 22$ counts
per bin) background flare 
(e.g. at time $\sim 1.83$ and $\sim 1.9$ in  
Figure~\ref{lmaraschi-C2_fig:unclean}), 
we conservatively excluded 3 more data points
(i.e. 3000 s) after the end of such flares.  The cleaned PN lightcurves
normalized to their mean are presented in Figure~\ref{lmaraschi-C2_fig:lc}
(colored points).

PKS2155-304 has been also observed with the OM using the UVW2 filter
($\sim 2000$ \AA ).  The OM was operated in the standard imaging mode,
with exposure time frames of 800~s, giving an almost continuous
coverage of the target.  The source counts were extracted in a
circular region with $r=48^{\prime \prime}$ and the background in a
concentric annular region from $48^{\prime \prime}$ to $2^{\prime
}$. The OM lightcurve normalized to the mean is shown in
Figure~\ref{lmaraschi-C2_fig:lc} (black points).

Some simultaneous ground based photometry and polarimetry is available 
for the second part of the observation and will be presented elsewhere.

\subsection{Temporal Analysis}

\begin{figure}[ht]
  \begin{center}
    \epsfig{file=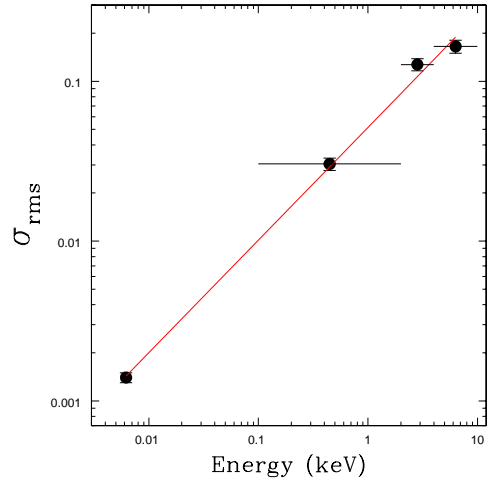, width=7cm}
  \end{center}
\caption{Normalized excess variance, $\sigma^2_{\rm rms}$, 
(as in Zhang et al. 2002) 
dependence on energy. The red line represents the best fit model.}
\label{lmaraschi-C2_fig:srms}
\end{figure}

To compare the amplitude of variability in the various energy bands we
calculated the normalized excess variance, $\sigma^2_{\rm rms}$ (Zhang et
al. 2002).  This is given in Figure~\ref{lmaraschi-C2_fig:srms}, which
clearly shows that the trend of increased variability with energy found
by e.g. Zhang et al. (2002), Edelson et al. (2001) in the X-ray regime
extends to the UV.
%$(0.14\pm0.01)10^{-2}$, $(3.04\pm0.27)10^{-2}$, 
%$(12.73\pm1.11)10^{-2}$, and $(16.54\pm1.53)10^{-2}$ for the optical,
%0.1--2.0~keV, 2--4~keV, and 4--10~keV light curve, respectively.
This trend is well modeled with a power law 
$\sigma^2_{\rm rms} = k \times (E/2.8284 {\rm keV})^{\gamma}$
with $\gamma = 0.70$ and $k=0.1070$.

\subsubsection{Cross-correlation}

To construct the cross-correlation function (CCF) we first normalized the
light curves to zero mean and unit variance by subtracting the mean count
rate and dividing by the rms (root mean squared) of the light curves
(e.g., Edelson et al. 2001).  The CCF is measured using the Discrete
Correlation Function (DCF, Edelson \& Krolik 1988). The bin size of the
DCF is chosen to be 1200~s. We first calculated the DCFs of the first
part of the observation which is the most variable during this campaign
(see Figure~\ref{lmaraschi-C2_fig:lc}). The DCF of the 0.1--2~keV versus
the 4--10~keV energy bands shows that variability in these two bands is
highly correlated, and, more important, there is no detectable time lag
between the two bands. This behaviour is confirmed by the DCF analysis
between other X-ray bands (i.e., 0.1--2 vs 2--4~keV, and 2--4 vs
4--10~keV).  The DCF of the OM versus any of the PN energy bands for the
first part of the observation shows an asymmetric structure at negative
lags of still unclear nature, which disappears considering the whole
observation.

%\begin{figure}[ht]
%  \begin{center}
%    \epsfig{file=DCF.eps, width=6cm}
%  \end{center}
%\caption{Discrete Correlation Function of the 0.1--2~keV versus
%the 4--10~keV PN energy bands}
%\label{lmaraschi-C2_fig:dcf}
%\end{figure}

\subsubsection{Structure Function}

\begin{figure}[ht]
  \begin{center}
    \epsfig{file=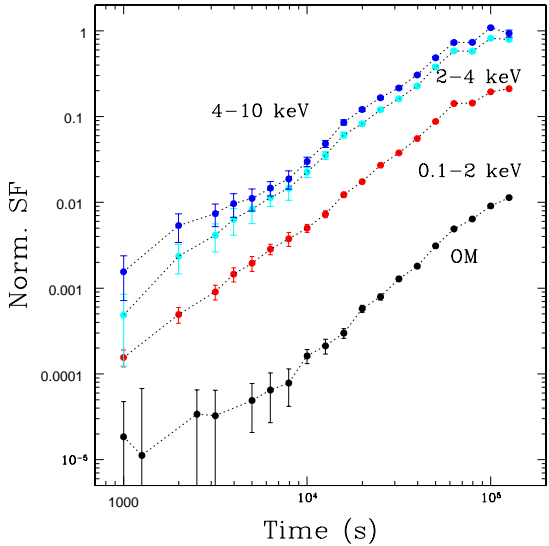, width=8cm}
  \end{center}
\caption{Normalized OM and PN (0.1--2~keV, 2--4~keV and 4--10~keV) 
structure functions (see text for details).}  
\label{lmaraschi-C2_fig:sf}
\end{figure}

We compute the Structure Function ($SF$) of each 1000~s binned lightcurve
normalized to the square of the lightcurve mean, as in Zhang et al. (2002), 
so that the $SF$s can be directly
compared.  We subtract from each $SF$ the contribution of the Poisson
noise.  The normalized $SF$s for the OM and for the 3 PN energy bands are
shown in Figure~\ref{lmaraschi-C2_fig:sf}.  As expected from the trend
clearly visible in Figure~\ref{lmaraschi-C2_fig:lc} and
Figure~\ref{lmaraschi-C2_fig:srms}, the amplitude of the variability
(i.e. the normalization of the $SF$ curve) increases with the energy of
the lightcurve.

We fit the normalized structure functions at times $> 10^4$~s 
(to exclude the region dominated by the Poisson noise) with a power
law model: $SF(\tau) = k \times (\tau/20 {\rm ks})^{\beta}$.
The best fit values for the normalization ($k$) and for the slope
($\beta$), together with their $1\sigma$ errors are reported in
Table~\ref{lmaraschi-C2_tab:tab1}.
A fit with a broken power law model suggests a break very close to the 
longest sampled timescales and does not improve the fit significantly. 

\begin{table}[bht]
  \caption{Fits to the normalized PN and OM  structure functions 
assuming a power law model.}
  \label{lmaraschi-C2_tab:tab1}
  \begin{center}
    \leavevmode
    \footnotesize
    \begin{tabular}[h]{lccc}
      \hline \\[-5pt]
      Energy band 	&$\beta$     	  &$k$$^a$   &$\chi ^2 / \nu$ \\[+5pt]
      \hline \\[-5pt]
      OM  		&$1.69 \pm 0.71$  &$0.620 \pm 0.006$	&8.0\\
      0.1--2 keV  	&$1.44 \pm 0.01$  &$19.4 \pm 1.4$	&26.6\\
      2--4 keV  	&$1.39 \pm 0.01$  &$86.8 \pm 0.7$	&16.8\\
      4--10 keV  	&$1.41 \pm 0.01$  &$113.7 \pm 10.8$	&12.2\\
      \hline \\
      \end{tabular}
  \end{center}
$^a$ In units of $10^{-3}$.\\
\end{table}

The large errorbars hide any possible dependence of the $SF$ slope on the
energy.

\subsection{Spectral analysis}

We divided the first part of the observation into two intervals:
the ``high'' and ``low'' flux states (see Figure~\ref{lmaraschi-C2_fig:lc}).

We present the first {\em XMM-Newton} EPIC spectrum of PKS2155-304.  Due
to unsolved problem of the calibration we restricted the analysis to the
0.6--10~keV energy range.  We analyzed the spectral data using XSPEC
(v.11) and the latest version of the response matrices (2001-12-11).

\begin{table}[bht]
  \caption{Spectral fit to the high and low states of the first 
part of the observation assuming a broken power law model.}
  \label{lmaraschi-C2_tab:tab2}
  \begin{center}
    \leavevmode
    \footnotesize
    \begin{tabular}[h]{lcccc}
      \hline \\[-5pt]
      & $\Gamma _1$&$\Gamma _2$  &$E_b$     &$\chi ^2 / \nu$ \\[+5pt]
      \hline \\[-5pt]
      High  &  $2.73 \pm0.01$ & $2.9 \pm 0.1$& $4.6\pm 1.0$  & 1.19\\
      Low   & $2.82 \pm 0.01$ & $2.97\pm 0.05$  & $3.6\pm0.5$ &1.46\\
      \hline \\
      \end{tabular}
  \end{center}
\end{table}

A simple (Galactic) absorbed power-law model is clearly inadequate to
reproduce the spectrum. We then used the broken power-law model, which
provides a good fit, except for an evident deficit of photons close to
$\sim 2.2$~keV.  This feature is located near to the Au instrumental
edge in the telescope effective area and indicates residual
uncertainties in the calibration at this energy in the SW mode.

The spectral indices measured in the high state ($\Gamma _1=2.73 \pm
0.01$ and $\Gamma _2=2.9 \pm 0.1$) indicate that the spectrum is
steepening with energy. During the low state the data indicate a further
significant steepening of the spectrum $\Gamma _1=2.82 \pm 0.01$ and
$\Gamma _2=2.97 \pm 0.05$, consistent with what is usually observed in BL
Lacertae objects (Urry et al. 1997; Kataoka et al. 2000). The 2--10
keV fluxes for the two states are $2.8\times 10^{-11}$ erg cm$^{-2}$
s$^{-1}$ and $2.0 3\times 10^{-11}$ erg cm$^{-2}$ s$^{-1}$, respectively.
PKS2155-304 is thus in a ``low'' state, similar to the 1996 {\it
Beppo-SAX} observation (Giommi et al. 1998; Zhang et al. 2002).

The curved model of Fossati et al. (2000) does not improve the fit and t
is not considered here.

\begin{figure}[ht]
  \begin{center}
    \epsfig{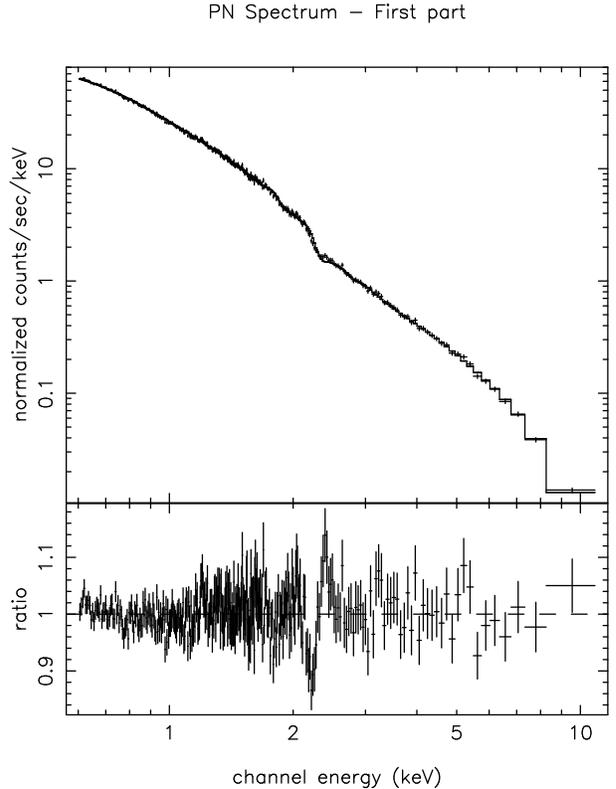}
  \end{center}
\caption{Energy spectrum of high state of the first part of the observation 
(top panel) and residuals to the fit (bottom panel). 
The feature at $\sim 2$~keV is likely to be related to calibration 
uncertainties.}
\label{lmaraschi-C2_fig:spec}
\end{figure}

\begin{figure}[ht]
  \begin{center}
    \epsfig{file=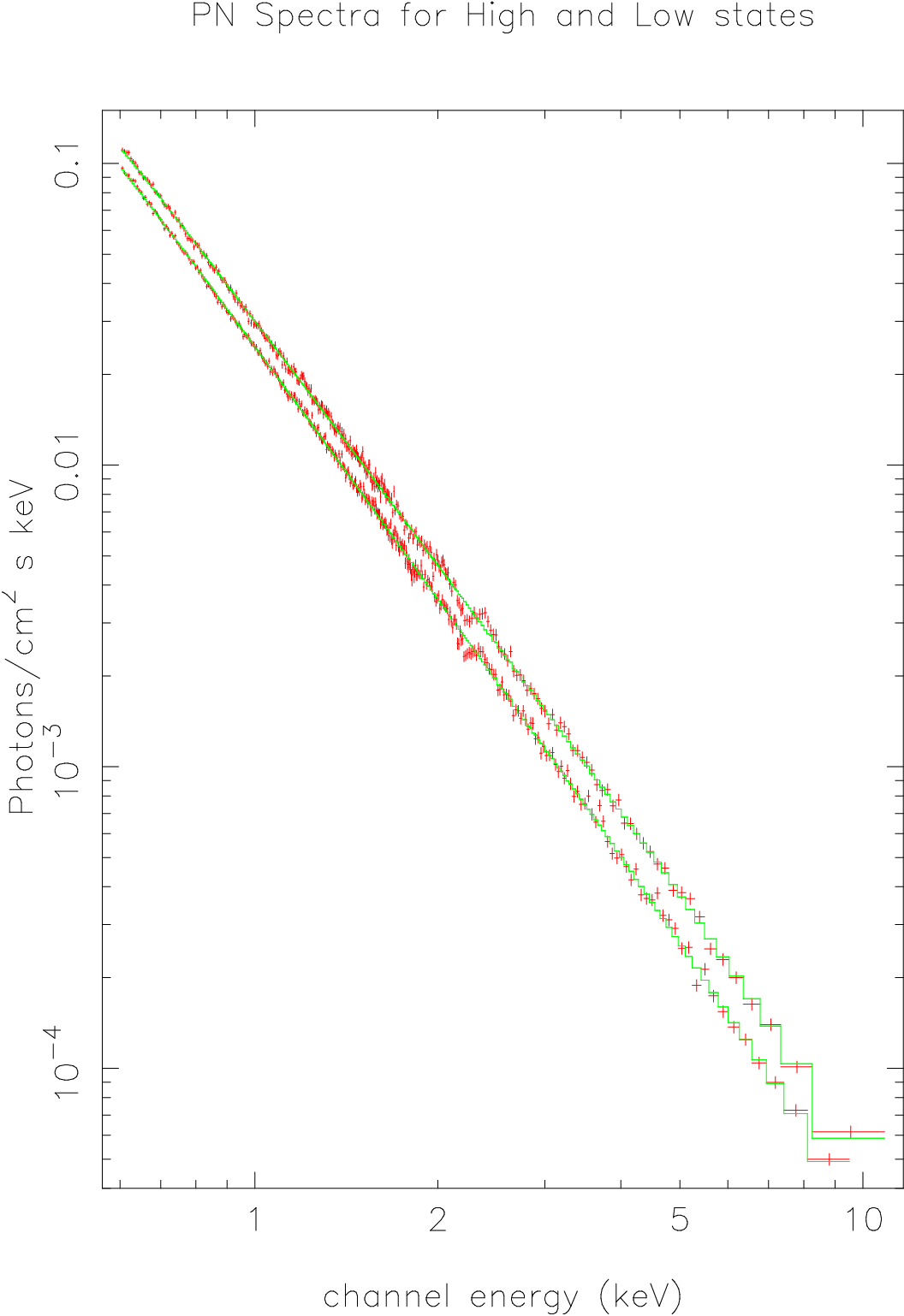, width=8cm}
  \end{center}
\caption{Unfolded spectra of high and low states of the first part of the
the observation.}
\label{lmaraschi-C2_fig:ufspec}
\end{figure}

\section{Discussion}

During the observation the source was monotonically dimming with some
superposed flickering. The most remarkable aspect is the energy
dependence of the dimming, which is a factor 3 in 4--10~keV range and
20\% in the UV (Figure~\ref{lmaraschi-C2_fig:lc} and
\ref{lmaraschi-C2_fig:srms}).  Figure~\ref{lmaraschi-C2_fig:srms} shows
that variability increases regularly with the energy. Moreover the
structure functions indicate that the dependence with the energy holds at
all the timescales. Both results are consistent with the results of Edelson
et al. (2001) and Zhang et al. (2002), but here are rather apparent, and
are shown to extend to the UV range.

It is noticeable that in the presence of such a clear variability/energy
dependence, the shape of the structure functions do not differ within the
errors, indicating that the same variability mechanism is operating at
all the investigated energies.\\

A further important point is the finding of no lags between different
X-ray ranges. This is consistent with what reported by Edelson et al.
(2001) in their {\em XMM-Newton} observation of PKS 2155-304, but it is
an apparent disagreement with previous findings of time lags by the {\em
ASCA} and {\em BeppoSAX} satellites (Kataoka et al. 2001; Tanihata et
al. 2001; Zhang et al. 2002). It should be noted, however, that time lags
have been found only in high-flux states, while both {\em XMM-Newton}
observations found the source in a low state. Moreover, while during {\em
ASCA} and {\em BeppoSAX} PKS 2155-304 was highly active and lightcurves
included well defined flares, the {\em XMM-Newton} lightcurves are rather
smooth, showing only slow variations. These differences could suggest
that {\it variability is driven by different mechanisms in the low and
the high state}. In order to assess this proposal, further observations
in the X-ray/$\gamma $-ry domain are necessary.

\begin{acknowledgements}
We thank Valentina Braito for valuable help in the analysis of XMM
data. We acknowledge support from Italian MUIR (contract COFIN
MM02C71842) and ASI (grant I-R-105-00).
\end{acknowledgements}


\begin{thebibliography}{}

%\bibitem[\protect\astroncite{Allen}{1973}]{fauthor-E1:all73}
%Allen C. 1973, Astrophysical quantities, Athlone Press

\bibitem[Edelson \& Krolik(1988)]{1988ApJ...333..646E} Edelson, R.~A.~\&
Krolik, J.~H.\ 1988,  ApJ, 333, 646

\bibitem[Edelson et al.(2001)]{2001ApJ...554..274E} Edelson, R.,
Griffiths, G., Markowitz, A., Sembay, S., Turner, M.~J.~L., \& Warwick,
R.\ 2001,  ApJ, 554, 274

\bibitem[Fossati et al.(2000)]{2000ApJ...541..166F} Fossati, G.~et al.\
2000, ApJ, 541, 166 

\bibitem[Giommi et al.(1998)]{1998A&A...333L...5G} Giommi, P.~et al.\
1998, A\&A, 333, L5

\bibitem[Kataoka et al.(2000)]{2000ApJ...528..243K} Kataoka, J.,
Takahashi, T., Makino, F., Inoue, S., Madejski, G.~M., Tashiro, M., Urry,
C.~M., \& Kubo, H.\ 2000, ApJ, 528, 243 

\bibitem[Kataoka et al.(2001)]{2001ApJ...560..659K} Kataoka, J.~et al.\
2001, ApJ, 560, 659 

\bibitem[Nicastro et al.(2002)]{} Nicastro, F. et al. 2002, ApJ, in press
(astro-ph/0201058).

\bibitem[Sikora \& Madejski(2001)]{2001hegr.proc..275S} Sikora, M.~\&
Madejski, G.\ 2001, High Energy Gamma-Ray Astronomy, 275

\bibitem[Tanihata et al.(2001)]{2001ApJ...563..569T} Tanihata, C., Urry,
C.~M., Takahashi, T., Kataoka, J., Wagner, S.~J., Madejski, G.~M.,
Tashiro, M., \& Kouda, M.\ 2001,  ApJ, 563, 569

\bibitem[Tavecchio, Maraschi, \& Ghisellini(1998)]{1998ApJ...509..608T}
Tavecchio, F., Maraschi, L., \& Ghisellini, G.\ 1998, ApJ, 509, 608

\bibitem[Urry \& Padovani(1995)]{1995PASP..107..803U} Urry, C.~M.~\& Padovani, P.\ 1995, PASP, 107, 803

\bibitem[Urry et al.(1997)]{1997ApJ...486..799U} Urry, C.~M.~et al.\
1997, ApJ, 486, 799 

\bibitem[Zhang et al. (2002)]{} Zhang, Y.H., et al. 2002, ApJ, in press

\end{thebibliography}
\end{document}